\documentclass[twoside]{article}

\usepackage{fleqn,espcrc2, epsfig}

\newcommand{\sla}[1]{#1\!\!\!/}

\title{Hadronic Decays of Excited Heavy Mesons}
\author{M. Di Pierro\thanks{Talk presented by} and E. Eichten \\
Fermilab, Batavia, IL 60510, USA}

\begin{document}

\begin{abstract}
We studied the hadronic decays of excited states of
heavy mesons ($D$, $D_s$, $B$ and $B_s$) to lighter states by emission 
of $\pi$, $\eta$ or $K$.
Wavefunctions and energy levels of these excited states are determined 
using a Dirac equation for the light quark in the potential generated by the heavy quark 
(including first order corrections in the heavy  quark expansion). 
Transition amplitudes are computed in the context of the Heavy Chiral Quark Model.
\end{abstract}
\maketitle

\section{INTRODUCTION}

In the context of the most general Quark Model, a heavy-light meson ($H$) is modeled with a light quark ($q$) bound to a static source of chromo-electro-magnetic field (the heavy quark $h$). Its hadronic transitions can be computed assuming that only the light quark enters in the reaction through and effective coupling $g_A$ (Heavy Chiral Quark Model~\cite{quarkmodel}).

Our work generally follows that of ref.~\cite{isgur,eichten,roberts,goity}. We differ in the choice of the potential. Moreover, we included for the first time the mixing effects in the spectrum and decay amplitudes.

\section{SPECTRUM}

\subsection{Notation}

The Dirac wavefunction, $\Psi$, of the light quark can be determined by solving 
the eigenvalue problem
\begin{equation}
H \Psi_{n,\ell,j,J,M} = E_{n,\ell,j,J} \Psi_{n,\ell,j,J,M}
\label{dirac}
\end{equation}
where $H$ is the Hamiltonian of the system and $\Psi$ is 4-spinor that represents the wavefunction of the system. In our notation
\begin{equation}
\Psi_{n \ell j J M} = C^{J,M}_{j,m;\frac12,S} \xi_S \left( 
\!\!\!\!\!
\begin{tabular}{rll}
$i f^0_{n \ell j}$ & $\!\!\!\!\! k^{+}_{\ell,j,m}$    & $\!\!\!\!\! Y^{\ell}_{m-\frac12}$ \\
$i f^0_{n \ell j}$ & $\!\!\!\!\! k^{-}_{\ell,j,m}$    & $\!\!\!\!\! Y^{\ell}_{m+\frac12}$ \\
$f^1_{n \ell j}$   & $\!\!\!\!\! k^{+}_{2j-\ell,j,m}$ & $\!\!\!\!\! Y^{2j-\ell}_{m-\frac12}$ \\
$f^1_{n \ell j}$   & $\!\!\!\!\! k^{-}_{2j-\ell,j,m}$ & $\!\!\!\!\! Y^{2j-\ell}_{m+\frac12}$
\end{tabular} \!\!\!\!
\right) 
\label{4spinor}
\end{equation}
where $n$ is the radial quantum number, $\ell$ is the orbital quantum number, $j,m$ are the total spin of the light quark and its $z$ component, $J,M$ are the total spin of the meson and its $z$ component, $f^0$,$f^1$ are radial wavefunctions and $Y^\ell_m$ are the usual spherical harmonics. $\xi_S$ is the 2-spinor associated to the heavy quark $h$ and $S$ is its spin.

Our convention for the phase and the normalization is such that
\begin{equation}
k^{\pm}_{\ell,j,m} = \Bigg\{
\begin{tabular}{ll}
$\sqrt{\frac{\ell \pm m+\frac12}{2\ell+1}}$
& if $j=\ell+\frac12$ \\
$\pm \sqrt{\frac{\ell \mp m+\frac12}{2\ell+1}}$ 
& if $j=\ell-\frac12$
\end{tabular} 
\end{equation}

\subsection{Choice of the potential}

Ignoring $1/m_h$ corrections, the most general form of the Hamiltonian that appear in eq.~(\ref{dirac}) is
\begin{equation}
H^{(0)} = -i \gamma^0 \gamma^i \partial_i + \gamma^0 m_q + \gamma^0 V_s + M_h + V_v
\label{dirac0}
\end{equation}
where $V_s$ is a spin independent potential, $V_v$ is a spin dependent potential and $M_h$ is a total energy shift (not to be confused with the mass of the heavy quark $m_h$ that appear in corrections to $H=H^{(0)}+O(1/m_h)$).

Asymptotic freedom suggests that at short distances $V_v \simeq 1/r$ dominates, while lattice simulations suggest that at large distances $V_s \simeq r$ dominates. As it was observed in ref.~\cite{isgur}, the choice of Coulomb-like potential at short distance is inconsistent with $1/m_h$ spin-dependent correction to $H^0$, because of ultraviolet divergences. The solution of the problem is that, in the context of the Dirac equation with a finite $m_h$, it is not correct to localize the heavy quark with a delta function since one must take into account the spatial degrees of freedom of the heavy quark. Our pragmatic approach to the problem is that of delocalizing the heavy quark within a length scale $1/\lambda$ assuming a Gaussian wave-function, $\Phi(x)$, for the former. The effective potential felt by the light quark is, therefore, a convolution of the Coulomb-like potential with the square of the wave-function of the quark:
\begin{equation}
V_v(r)=\int  |\Phi({\mathbf x})|^2 \frac{\alpha_s}{|{\mathbf r} - {\mathbf x}|}\textrm{d}^3{\mathbf x} = \frac{\alpha_s}{r}\textrm{erf}(\lambda r) 
\label{vv}
\end{equation}
Our choice for the spin-independent part of the potential is
\begin{equation}
V_s(r) = b r  + c
\label{vs}
\end{equation}
(notice from eq.~(\ref{dirac0}) that $c$ is not a physical parameter since it can be re-absorbed into the definition of $m_q$).

\begin{figure}
\begin{center}
\epsfxsize=7.5cm
\epsfysize=7.5cm
\epsfbox{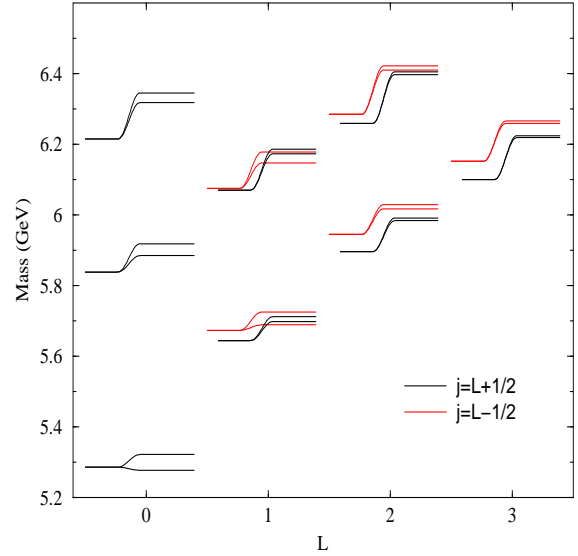} 
\end{center}
\vskip -1.5cm
\caption{Spectrum of excited B mesons.\label{spectrum}}
\vskip 1cm
\begin{center}
\begin{tabular}{lllrr} \hline
Particle & $n^j\ell_J$ & Exp. & Model & $\Gamma_{\textrm{tot}}/g_A^2$ \\ \hline
$D$	         & $1^\frac12S_0$	& $1864$	& 	$1871$	&  $-$ \\ 
$D^\ast$	 & $1^\frac12S_1$	& $2007$	& 	$2006$	&  $-$ \\ 
$D_1$	         & $1^\frac32P_1$	& $2427$	& 	$2420$	&  $25.0$ \\ 
$D^\ast_2$	 & $1^\frac32P_2$	& $2459$	& 	$2462$	&  $38.2$ \\ \hline

$D_s$	& $1^\frac12S_0$		& $1969$	& 	$1965$	&  $-$ \\
$D_s^\ast$	& $1^\frac12S_1$	& $2114$	& 	$2112$  &  $-$ \\ 
$D_{s1}$	& $1^\frac32P_1$	& $2535$	& 	$2535$	&  $94.3$ \\ 
$D^\ast_{s2}$	& $1^\frac32P_2$	& $2573$	& 	$2579$	&  $4.9$ \\ \hline

$B$		& $1^\frac12S_0$	& $5279$	& 	$5278$	&  $-$ \\ 
$B^\ast$	& $1^\frac12S_1$	& $5325$	& 	$5322$	&  $-$ \\ 
$B_0$		& $1^\frac12P_0$	& 		& 	$5689$	&  $174.6$ \\ 
$B_1$		& $1^\frac32P_1$	& 		& 	$5698$	&  $14.3$ \\ 
$B^\ast_1$	& $1^\frac12P_1$	&   		&       $5725$	&  $168.0$ \\
$B^\ast_2$	& $1^\frac32P_2$	&  		& 	$5712$	&  $20.0$ \\
$B^{\prime}$	 & $2^\frac12S_0$	& 		& 	$5885$	&  $28.1$ \\ 
$B^{\prime\ast}$ & $2^\frac12S_1$	& 		& 	$5918$	&  $49.3$ \\ \hline 

$B_s$	& $1^\frac12S_0$		& $5369$	& 	$5369$	&  $-$ \\ 
$B_s^\ast$	& $1^\frac12S_1$	& $5416$	& 	$5417$	&  $-$ \\ 
$B_{s1}$	& $1^\frac32P_1$	& 		& 	$5801$	&  $-$ \\ 
$B^\ast_{s2}$	& $1^\frac32P_2$	& 		& 	$5815$	&  $0.2$ \\ \hline
\end{tabular}
\end{center}
Table 1. Tabulated spectrum for the observed states used in the fit together 
with predictions for some excited $B$ states. All units are in MeV.
\end{figure}

\subsection{$1/m_h$ correction}

We solve the eigenvalue problem associated to the eq.~(\ref{dirac}) using the Hamiltonian in eq.~(\ref{dirac0}) and the potentials~(\ref{vv},\ref{vs}). In this way we determine the radial wave-functions $f^0_{n \ell}$, $f^1_{n \ell j}$ and the associated eigenvalues $E_{n \ell j}$.

$1/m_h$ corrections to the the Hamiltonian have been derived in \cite{roberts} in the Bethe-Salpeter formalism. They are responsible for the spin-orbit interaction, the hyperfine spitting and the mixing of states with the same $j$. We include these effects as a perturbation to the energy levels ($\delta E _{n \ell j J}$) and also determine the mixing coefficients for each doublet of states.

In terms of the $f^i$ the $1/m_h$ correction to the energy levels reads as
\begin{equation}
\delta E = \int (A+B+C) r^2 \textrm{d}r
\end{equation} 
where one can identify the contributions proportional to $p^2$:
\begin{equation}
A = -\sum _i f^i \left(\partial^2_r + \frac2r \partial_r - \frac{\ell^2+\ell}{r^2}\right) f^i
\end{equation}
the spin orbit interaction, for $ j=\ell + \frac12$:
\begin{equation}
B= V_v \!\! \left[ f^1 (\partial_r -\frac{\ell}r )f^0 \!\!-\!\! f^0 (\partial_r  + \frac{\ell+2}r )f^1 \right]
\end{equation}
for $j=\ell - \frac12$:
\begin{equation}
B= V_v \!\! \left[ f^1 (\partial_r \! + \! \frac{\ell+1}r )f^0 \!\! - \!\! f^0 (\partial_r \! - \! \frac{\ell-1}r )f^1 \right]
\end{equation}
and the hyperfine splitting contribution:
\begin{equation}
C = (-1)^{J-j} \frac{2j+1}{2J+1} V_v' f^0 f^1
\end{equation}

\begin{figure}
\begin{center}
\epsfxsize=7.5cm
\epsfysize=7.5cm
\epsfbox{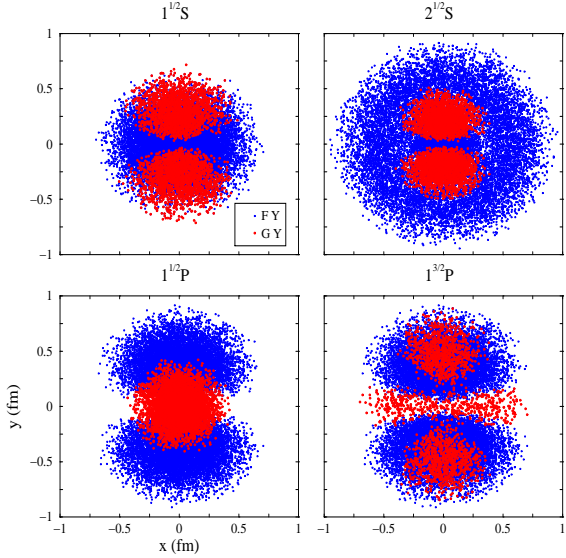} 
\end{center}
\vskip -1cm
\caption{Orbitals for some excited B mesons.\label{density}}
\end{figure}

\subsection{Choice of parameters and predictions}

The nine parameters of the model ($\alpha_s$, $\lambda$, $b$, $m_q|_{q=u}$, $m_q|_{q=s}$, $m_h|_{h=c}$, $M_h|_{h=c}$, $m_h|_{h=b}$, $M_h|_{h=b}$) are determined by best fit (minimum $\chi^2$) to the known spectrum of excited states of the $D$, $D_s$, $B$ and $B_s$ mesons. Our results are:
\hskip -1cm
\begin{center}
\begin{tabular}{ll}
$\alpha_s = 0.339$ & \\
$\lambda = 2.820$ \textrm{GeV}& \\
$b = 0.257$ \textrm{GeV} & \\
$m_u = 0.073$ \textrm{GeV} & $m_s = 0.214$ \textrm{GeV} \\
$m_c = 1.52$ \textrm{GeV} & $M_c = 1.52$ \textrm{GeV} \\
$m_b = 4.67$ \textrm{GeV} & $M_b = 4.68$ \textrm{GeV} 
\end{tabular}
\end{center}
\hskip -1cm

The spectrum for some of the states plotted in fig.~\ref{spectrum} and tabulated in table~1.
A density plot of $|f^0_{n \ell j} Y^{\ell}_0|^2$ and $|f^1_{n \ell j} Y^{2j-\ell}_0|^2$ for some excited states of a B meson is reported in fig.~\ref{density}.

\subsection{Tests of the model}

\begin{figure}
\begin{center}
\epsfxsize=7.5cm
\epsfysize=7.5cm
\epsfbox{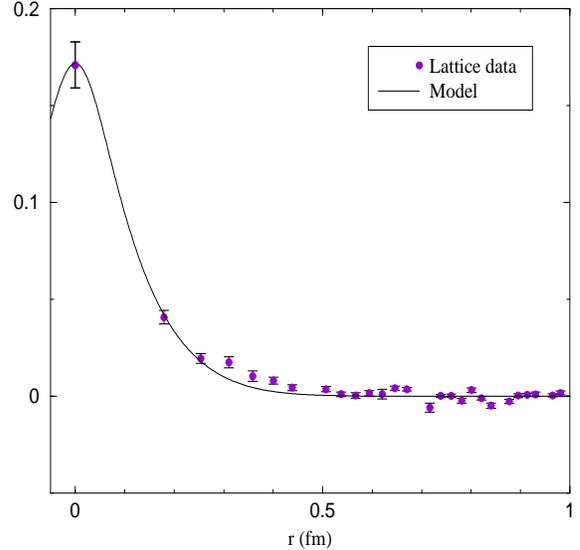} 
\end{center}
\vskip -1cm
\caption{Comparison between a prediction of our model and the lattice QCD result~\cite{me}.\label{axial}}
\end{figure}

Despite the good fit of the mass spectrum we decided to test our model by comparing some of the transition amplitudes with some recent lattice results. In particular we computed
\begin{equation}
{\mathcal A}_{B^\ast \rightarrow B \pi}(r) = \int \left< B^\ast \right| A_\mu({\mathbf r}) \left| B \right> \textrm{d} \Omega_r
\end{equation}
using our chiral quark model (where the only unknown parameter is the overall normalization, $g_A$, which is the effective coupling of the quark to the axial current, $A_\mu$) and comparing it with the lattice result of ref.~\cite{me}. The comparison is shown in fig.~\ref{axial}. In the plot
the point at $r=0$ is used to fix the relative normalization. A more sophisticated analysis 
of the lattice results gives%
\footnote{this is a preliminary determination valid in the limit $p_\pi,m_\pi \rightarrow 0$} 
$g_A=0.42 \pm 0.09$.

\section{HADRONIC DECAYS}

\subsection{Decay amplitudes}

We consider here the most general hadronic transition
\begin{equation}
H' \rightarrow H + x
\label{transition}
\end{equation}
where $H'$ is an heavy meson with associated wave-function $\Psi'$, $H$ is an heavy meson with associated wave-function $\Psi$ and $x$ is a light meson with momentum ${\mathbf p}$. Such a transition is mediated by a matrix element of the form
\begin{equation}
I(X, {\mathbf p}) = g_A \int \bar \Psi^\ast X e^{-i {\mathbf p} {\mathbf r}} \Psi' \textrm{d}^3r
\end{equation}
where $X$ is the $4\times4$ spin matrix that characterize the transition.
In the particular case in which the light meson $x$ is a pseudoscalar ($\pi$, $\eta$ or $K$) 
$X=\sla p \gamma^5$, while if the light meson $x$ is a vector ($\rho$, $\omega$ or $K^\ast$)
$X=\sla \epsilon$ (and $\epsilon$ is the polarization vector of the outcoming vector meson).

The exponential can be expanded in products of spherical harmonics and spherical 
Bessel functions ($j_k$), thus giving
\begin{equation}
I(X, {\mathbf p}) = \!\! \sum_{\ell_x, m_x} Y^{\ell_x \, \ast}_{m_x}(\hat p)
C^{J',M'}_{J,M;\ell_x,m_x} {\mathcal A}_{\ell_x}(X,p) 
\end{equation}
We computed ${\mathcal A}_{\ell_x}(X,p)$ for a complete set of spin matrices $X$ and proved that 
it can alway be reduced to integrals of the form
\begin{equation}
{\mathcal A}_{\ell_x}(X,p) = g_A \!\!\! \sum_{i,j=0,1}\!\! c^{ijk}_{\ell_x}(X) \int_0^\infty \!\!\! (f^i j_k f^j) r^2 \textrm{d}r
\end{equation}
where $c^{ijk}_{\ell_x}(X)$ depend also on the quantum numbers of the mother and the daughter mesons but not on the radial wavefunctions. 

\subsection{Decay widths}

The decay width for transition of eq.~(\ref{transition}), when light meson $x$ is emitted in an eigenstate of the total momentum $p$ and of the angular momentum $\ell_x$, is given by
\begin{equation}
\Gamma_{\ell_x}^{H'Hx} = \frac{\zeta^2}{16 \pi^2 f_\pi^2} \frac{2 J +1}{2 J' +1} \frac{M}{M'} \left|   {\mathcal A}_{\ell_x}(\sla p \gamma^5,p) \right|^2
\end{equation}
where $\zeta=\sqrt{3}$, $1/\sqrt{3}$, $2/\sqrt{3}$ or $1$ for $\pi$, $\eta$ (for a nonstrange heavy meson), $\eta$ (for a strange heavy meson) or $K$ respectively; $M',J'$ and the mass and total angular momentum of the mother meson $H'$; $M,J$ are the mass and angular momentum of the daughter meson $H$, and $p$ can be determined by energy-momentum conservation ($p_0=M'-M$, $p=|{\mathbf p}|=\sqrt{p_0^2-m_x^2}$). The total decay width and the branching ratios are defined as
\begin{equation}
\Gamma_{\textrm{tot}}^{H'} = \!\sum_{x,\ell_x} \Gamma_{\ell_x}^{H'Hx}; \hskip 1mm 
Br(H'\!\!\rightarrow\!\!Hx;\ell_x) = \! \frac{\Gamma_{\ell_x}^{H'Hx}}{\Gamma_{\textrm{tot}}^{H'}}
\end{equation}
where the sum on $x$ spans all the hadronic decay modes with emission of a light pseudoscalar meson. The total width for some of the mesons is reported in table~1.

\section{CONCLUSIONS}

We computed the spectrum and the width of hadronic decays of excited $D$, $D_s$, $B$ and $B_s$ mesons in the context of the chiral quark model. As an example, we report here some hadronic decay channels for the first radial excited B meson (not including the $\rho$ decays)
\begin{center}
\begin{tabular}{lll} 
$B(2^\frac12S_0)$ & & \\
\hskip 5mm $\hookrightarrow B(1^\frac12S_1) + \pi$ & $\!\!\! (\ell_\pi=1$ & $\!\!\! Br = 77\%)$ \\
\hskip 5mm $\hookrightarrow B(1^\frac12P_0) + \pi$ & $\!\!\! (\ell_\pi=0$ & $\!\!\! Br = 22\%)$ \\
\hskip 10mm $\hookrightarrow B(1^\frac12S_0) + \pi$ & $\!\!\! (\ell_\pi=0$ & $\!\!\! Br = 100\%)$ \\
\hskip 5mm $\hookrightarrow B(1^\frac32P_2) + \pi$ & $\!\!\! (\ell_\pi=2$ & $\!\!\! Br = 0.33\%)$ \\
\hskip 5mm $\hookrightarrow B(1^\frac12S_1) + \eta$ & $\!\!\! (\ell_\eta=1$ & $\!\!\! Br = 0.03\%)$ 
\end{tabular}
\end{center}

More complete tables will be published on a separate paper.

We finally remark how our model is able to fit the masses of observed excited states within less than 10 MeV discrepancy (within 4 MeV in average) better than was done in preceding works.
\vskip 5mm
This work was performed at Fermilab, a U.S. Department of Energy Lab (operated by the University Research Association, Inc.), under contract DE-AC02-76CHO3000.

\end{document}